\documentclass[10pt,prl,aps,twocolumn,floatfix,showpacs]{revtex4}
\usepackage{graphicx}
\usepackage{amssymb}
\usepackage{bm}

\begin{document}

\title{Scale-renormalized matrix-product states for correlated quantum systems}

\author{Anders W. Sandvik} 
\affiliation{Department of Physics, Boston University, 590 Commonwealth Avenue, Boston, Massachusetts 02215}
\affiliation{Department of Physics, National Taiwan University, Taipei, Taiwan 106}

\date{\today}

\begin{abstract}
A generalization of matrix product states (MPS) is introduced which is suitable for describing interacting
quantum systems in two and three dimensions. These {\it scale-renormalized matrix-product states} (SR-MPS) are
based on a course-graining of the lattice in which the blocks at each level are associated with matrix products
that are further transformed (scale renormalized) with other matrices before they are assembled to form blocks at the next 
level. Using variational Monte Carlo simulations of the two-dimensional transverse-field Ising model as a test, it is shown 
that the SR-MPS converge much more rapidly with the matrix size than a standard MPS. It is also shown
that the use of lattice-symmetries speeds up the convergence very significantly.

\end{abstract}


\maketitle

In a variational study of an interacting quantum system, a wave function $\Psi$ with a number of adjustable 
parameters $p_1,\ldots,p_m$ is optimized by minimizing its energy $E=\langle \Psi |H|\Psi\rangle$ with respect 
to a hamiltonian $H$. Ideally, one would like to consider a functional form which allows for a systematic way of 
improving the calculation, by increasing the number of parameters $m$ in such a way that $\Psi$ is guaranteed to 
approach the true ground state of $H$ in the limit $m \to \infty$. 
A trivial way, in principle, is to expand $|\Psi\rangle$ in a complete set of states; $|\Psi\rangle = \sum_n c_n |n\rangle$, 
whence the parameters to be optimized are the wave function coefficients $c_n$ themselves. However, in practice, the 
Hilbert space is too large ($2^N$ states in the simplest case of $N$ spins with $S=1/2$) to include all states, and in 
general there is no obvious way to order the states so that their contributions to the ground state decrease 
as a function of $n$. To achieve this, one can attempt to optimize the basis in some way, with the goal of obtaining a 
hierarchy of basis states which systematically and rapidly improve the result as they are included in the calculation. 
This is the basic idea of renormalization group (RG) methods, which after decades of attempts, following Wilson's pioneering 
solution of Kondo impurity problem \cite{wilson}, led to a break-through in the form of White's density matrix RG (DMRG) 
method \cite{white,schollwock} for one-dimensional systems. 

For systems in higher dimensions, there has been recent progress in generalizing the DMRG approach, which is closely 
related to matrix-product states (MPS) \cite{ostlund}, using tensor-network states \cite{nishino}, e.g., the projected 
entangled pair states (PEPS) \cite{peps} and related MPS-like string states \cite{string}, as well as schemes based
on entanglement renormalization \cite{vidal}. However, there are still 
considerable challenges related to the convergence properties and computational complexity of these methods. 
In this Letter, an alternative class of generic correlated states---{\it scale-renormalized matrix-product states} 
(SR-MPS)---is introduced. These states combine concepts of coarse graining, renormalization, and MPS into a framework 
for systematically refined variational calculations. The scheme is tested on the two-dimensional transverse-field 
Ising model, using a recently developed variational Monte Carlo method \cite{awsandvidal} to optimize the SR-MPS.

In the DMRG method, the basis is re-optimized and truncated at some number $D$ of states as more sites are added to the 
lattice \cite{white}. Originally this was not viewed as a variational method, but it was soon recognized that the DMRG in 
effect produces the best variational MPS with $D\times D$ matrices \cite{ostlund}. For a system of $N$ spins represented 
by Pauli operators $\vec {\sigma}_i$, and working in the basis where all $\sigma^z_i$ are diagonal, $\sigma^z_i = \pm 1$, 
MPS for a periodic chain are of the form (using the notation $[ \sigma ]$ for $[\sigma^z_1,\ldots,\sigma^z_N]$)
\begin{equation}
|\Psi\rangle = \sum_{[\sigma^z]}W([\sigma ])|\sigma^z_1,\ldots,\sigma^z_N\rangle,
\end{equation}
where the wave-function coefficient is
\begin{equation}
W([\sigma ])={\rm Tr}\{ A(\sigma^z_1)A(\sigma^z_2)\cdots A(\sigma^z_N)\},
\label{wmps}
\end{equation}
and $A(\pm 1)$ are two $D \times D$ matrices. For systems with open boundaries, for which DMRG and standard MPS 
techniques are best suited in practice, the matrices are site dependent, and in stead of taking a trace the edge 
matrices are vectors. The DMRG method does not operate with 
MPS explicitly, but recently schemes have been devised for working directly with the MPS without invoking 
the DMRG procedures. This formally reduces the scaling of the computational effort from $D^6$ to $D^5$ for 
periodic systems \cite{verst}. Using Monte Carlo sampling of the spins states, instead of evaluating their
traces exactly, the scaling can be further reduced to $D^3$ \cite{awsandvidal,string}.

In higher dimensions, the DMRG method typically is implemented by regarding the system as a chain folded up to 
form the lattice of interest \cite{ladders}. In this effective one-dimensional system there are long-range interactions.
Further studies of the MPS formalism also showed why the DMRG method performs poorly in this case. The exponential 
scaling in the number of states that has to be kept \cite{liang}, or, equivalently, the matrix dimension $D$ of the MPS, 
is a consequence of the inability of the matrix products to account for entanglement between neighboring sites when the 
corresponding matrices are far apart \cite{vidal1}. To circumvent this problem, tensor-network states have been proposed 
as natural and effective generalizations of the MPS/DMRG to higher dimensions \cite{nishino,peps}.

\begin{figure}
\includegraphics[width=5.5cm, clip]{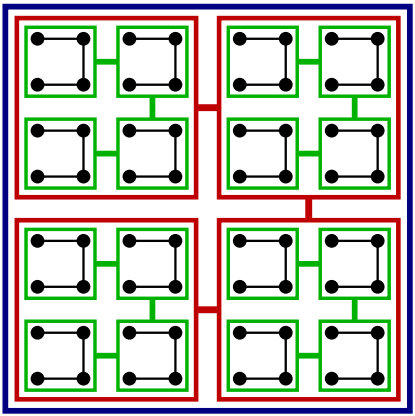}
\caption{(Color online) SR-MPS hierarchy for a square lattice. The circles represent the original matrices
$B^0_{x,y}=A(\sigma_{x,y})$. Products of four of these are indicated by connecting lines. A square enclosing such 
a unit represents scale renormalization with matrices $M^1_L,M^1_R$, which results in $B^1_{x,y}$. Successive 
levels of connected squares enclosed by larger squares represent products of four matrices $B^n_{x,y}$ followed by 
scale-renormalizations with $M^n_L,M^n_R$, resulting in $B^{n+1}_{x,y}$.}
\label{fig1}
\vskip-3mm
\end{figure}

Here SR-MPS is proposed as an alternative generalization of MPS for higher-dimensional systems. For a periodic 
system of $S=1/2$ spins $\vec {\sigma}_{x,y}$, each lattice site $(x,y)$, $x,y=1,\ldots, L$, is associated with a $D\times D$ 
matrix $A(\sigma^z_{x,y})$ as in the MPS. However, instead of just arranging these matrices according to a string on the 
lattice, the system is first subdivided into blocks, which are associated with matrix products. These matrix products are 
then {\it scale renormalized} by transforming them with some other matrices, before they are multiplied by similar 
block-matrices to represent a larger cell. For a square lattice, the resulting hierarchy of matrix products (in the 
simplest case based on blocks with four sub-blocks at each level) is illustrated in Fig.~\ref{fig1}. The block-matrices 
at levels $n$ and $n+1$ are related according to
\begin{equation}
B^{n+1}_{x,y}=M^n_L  B^n_{x,y}B^n_{x+1,y}B^n_{x+1,y+1}B^n_{x,y+1} M^n_R,
\label{bn1}
\end{equation}
with the lowest level corresponding to the original spin dependent matrices, $B^0_{x,y}=A(\sigma^z_{x,y})$, and $M^n_L,M^n_R$ 
accomplishing the scale renormalization. At level $n$ the block coordinates take the values $k2^n$, $k=1,\ldots, L/2^n$. Thus the 
lattice size should be a power of $2$; $L=2^l$. 

The purpose of the scale renormalization is to compensate, as much as possible, 
for the non-equivalent ways in which the four blocks at a given level $n$ are treated in the associated product of four matrices. 
It will be shown that the effect indeed is to make the four members of a block more uniform in their correlations with each other 
and the rest of the system.

Note that under the trace of the final assembly of products, $B^l_{1,1}$, an equivalent way of defining the block matrices
(\ref{bn1}) is with a single scale-renormalization matrix; $B^{n+1}_{x,y}=B^n_{x,y}B^n_{x+1,y}B^n_{x+1,y+1}B^n_{x,y+1} M^n$.
However, Eq.~(\ref{bn1}) allows for the possibility of increasing the matrix size with the level $n$, using rectangular
matrices $M^n_L$ and $M^n_R$ of size $D_{n+1} \times D_n$, and $D_n \times D_{n+1}$, respectively. This may be useful
if the scale-renormalization is further refined by making $M^n_{L,R}$ dependent on the physical state of the blocks
they transform, using, e.g., a block spin $\Sigma_{x,y} = 0,\pm 1$ [for six different matrices $M^n_{L,R}(\Sigma_{x,y})$]. 
Defining an appropriate block-variable for a given model is not always easy, however. In the 
Ising model considered here the Kadanov block-spin \cite{kadanov} can be used, but in this first 
study only the state-independent scale-renormalization (\ref{bn1}) will be applied and the matrix size will be 
kept constant; $D_n=D$. Keeping the symmetric form with left and right scale-renormalizations, instead of just a single 
$M^n$, seems to help in the optimization, in spite of the larger number parameters.

In the simplest version of the SR-MPS the wave function coefficient is the trace of 
$B^l\equiv B_{11}^l$. One can also use a sum of matrix products taken over symmetry transformations of the spin 
configuration. Spin-inversion symmetry can be used if the hamiltonian has it. Then
\begin{equation}
W([\sigma ])={\rm Tr}\{ B^l([\sigma ]) \pm B^l(-[\sigma ])\},
\label{ws}
\end{equation}
where $-[\sigma]$ denotes the configuration with all $\sigma_i \to -\sigma_i$. It is also useful to 
incorporate lattice symmetries. Denoting a transformation (including the identity) of $[\sigma]$ (translation, 
rotation, or reflection) by $T_r[\sigma]$, the wave function coefficient is, considering for simplicity a 
fully symmetric wave function with zero momentum,
\begin{equation}
W([\sigma ])=\sum_r {\rm Tr}\{ B^l(T_r[\sigma ]) + B^l(-T_r[\sigma ])\}.
\label{wrs}
\end{equation}
Here $r=1,\ldots,8N$ if all symmetries of the square lattice are used. It will be shown below that the use of symmetries
improves the $D$ convergence very significantly.

To test the SR-MPS scheme, it will be applied next to the Ising model in a transverse field;
\begin{equation}
H = -\sum_{x,y} (\sigma^z_{x,y} \sigma^z_{x+1,y}  + \sigma^z_{x,y} \sigma^z_{x,y+1} + h\sigma^x_{x,y}),
\end{equation}
with periodic boundaries ($\sigma^z_{L+1,y}=\sigma^z_{1,y}$ and $\sigma^z_{x,L+1}=\sigma^z_{x,1}$). Computations are 
expected to be the most challenging at quantum-critical points; $h$ in the vicinity of $h_c \approx 3.044$ \cite{hcrit} 
will be the main focus here. 
 
To optimize the wave function, here using general (non-symmetric) real matrices $A(\pm 1)$ and $M_L^n,M_R^n$, $n=1,\ldots, l$, 
the variational Monte Carlo method discussed in Ref.~\cite{awsandvidal} is used. The energy derivatives are calculated, and 
based on their signs the matrix elements are updated by a random amount, 
e.g., $a_{ij} \to a_{ij} -{\rm sign}(\partial E/\partial a_{ij})\delta r_{ij}$. Here $r_{ij} \in [0,1)$ is a 
random number and the maximum step $\delta$ is gradually reduced. If this reduction is sufficiently slow, the converged 
matrices will correspond to an energy minimum. Calculations in one-dimension \cite{awsandvidal} have shown that the global 
energy minimum, which evolves to the true ground-state energy with increasing $D$, can be reached with this method at
least up to $D \approx 50$

If no lattice symmetries are used, the Metropolis probability of flipping a spin can be evaluated with $\propto D^3$ 
operations using the sequential flip scheme of Ref.~\cite{awsandvidal}. However, with symmetries incorporated according 
to Eq.~(\ref{wrs}), the spins cannot be sequentially visited in all transformed configurations, and therefore a different 
scheme has to be employed. Organizing partial products in tree-structures, one for each lattice transformation, the total 
number of operations required for each spin update is $\propto N\ln(N)D^3$, where the factor $N$ is due to the number of
different matrix products and $\ln(N)$ comes from recalculating one branch of a tree.

\begin{figure}
\includegraphics[width=7.5cm, clip]{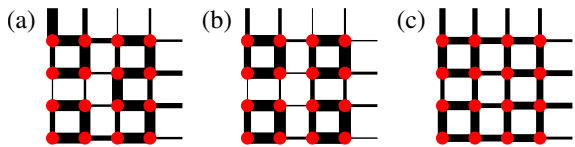}
\caption{(Color online) Nearest-neighbor spin correlations on a $4\times 4$ lattice at $h=3$ calculated using (a) 
an MPS with $D=2$, (b) an SR-MPS with $D=2$, and (c) an SR-MPS with $D=8$. The thinnest and thickest bars correspond, 
respectively, to a $-11\%$ and $+11\%$ deviation from the average. The bars at the right and upper edges represent 
the correlations across the boundaries (periodic boundary conditions are used). The energies $-E/N$ of the states in 
(a),(b),(c) are $3.1746$, $3.1772$, $3.2108$. The exact energy for $L=4$ is $-3.2155081$.}
\label{fig2}
\vskip-3mm
\end{figure}

First, an $L=4$ lattice will be considered. To demonstrate some of the effects of scale-renormalization, Fig.~\ref{fig2}
shows a plot of the spatial variations in the nearest-neighbor correlations $\langle \sigma^z_{x,y}\sigma^z_{x+1,y}\rangle$ 
and $\langle \sigma^z_{x,y}\sigma^z_{x,y+1}\rangle$, obtained with and without scale-renormalization. Spin-inversion 
symmetry is taken into account but no lattice symmetries are used, i.e., the states are sampled according to the
symmetric ($+$) Eq.~(\ref{ws}). 
With a small $D$ one would then expect to see traces of the particular way the blocks are constructed. Without 
scale-renormalization (i.e., $M^n_L,M^n_R=I$), the scheme reduces to an MPS calculation with the matrix product 
taken along the particular ``coarse-graining string'' used here in the SR-MPS. 
In Fig.~\ref{fig2}(a), obtained with $D=2$, it can be seen clearly that the correlations are non-uniform; in particular 
different sites within the $2\times 2$ blocks are not equally correlated with their neighbors. With scale-renormalization,
Fig.~\ref{fig2}(b) shows a significantly reduced non-uniformity between the blocks. The energy is also improved. With 
$D=8$ in the SR-MPS, shown in Fig.~\ref{fig2}(c), the non-uniformity is much reduced and the energy is improved 
considerably. In contrast, an MPS with $D=8$ (not shown) only marginally improves on the $D=2$ result.

\begin{figure}
\includegraphics[width=7.75cm, clip]{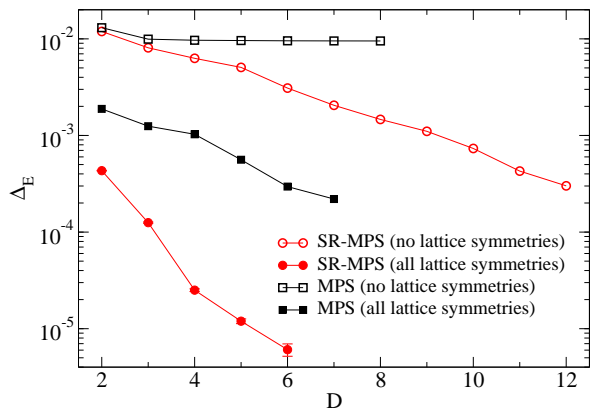}
\caption{(Color online) Relative deviation $\Delta_E=(E-E_D/E$ of the energy $E_D$ of SR-MPS and MPS with 
$D\times D$ matrices from the exact energy $E$ for a $4\times 4$ system at $h=3$.}
\label{fig3}
\vskip-3mm
\end{figure}

Fig.~\ref{fig3} illustrates the convergence of MPS and SR-MPS with an without lattice symmetries. 
Without lattice symmetries, the MPS energy converges extremely slowly and it is not possible in practice to obtain 
the ground state. By incorporating the lattice symmetries the convergence is substantially improved,
however. In the case of the SR-MPS, an exponential convergence with $D$ is seen even without lattice symmetries, and 
with these symmetries included the convergence is very rapid, with an accuracy of 
$10^{-5}$ reached already at $D=5$. Clearly, both lattice symmetries and scale renormalization have very favorable 
effects, and when using both of them the convergence properties seem very encouraging.

Before moving to larger lattices, a further improvement of the SR-MPS is noted: One can use different matrices 
for all the states of a block of spins, instead of just the two matrices $A(\sigma^z_{x,y})$ for individual spins. 
Using $2\times 2$ blocks, there are 16 matrices $A(\sigma^4_{x,y})$, where $\sigma^4_{x,y}=0,\ldots,15$ labels the states 
of the spins $\sigma^z_{x,y},\sigma^z_{x+1,y},\sigma^z_{x,y+1},\sigma^z_{x+1,y+1}$. This gives additional flexibility to the wave function, 
and thus a faster $D$ convergence can be expected. The larger number of matrices to optimize does not seem to pose any
difficulties in practice, and in addition roughly four times less operations are required for a spin update. 
Here calculations with $2\times 2$-block matrices $A(0,\ldots,15)$, will be compared with the basic $A(\pm 1)$ scheme.

\begin{figure}
\includegraphics[width=7.5cm, clip]{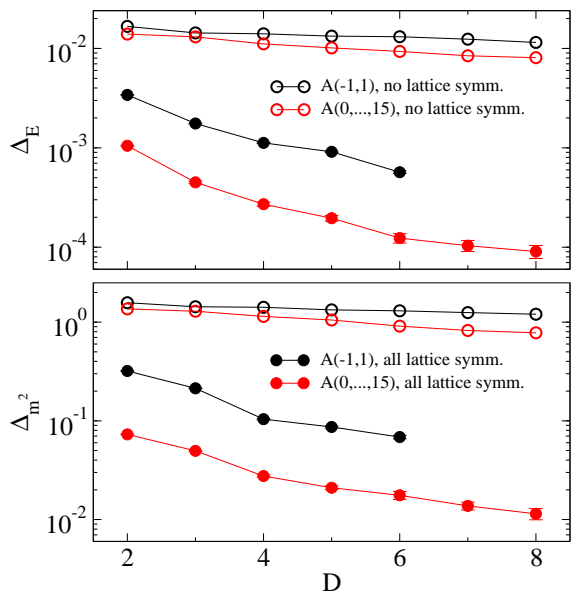}
\caption{(Color online) Relative error of the energy and the squared magnetization for an $L=8$ system at $h=3.044$, 
using SR-MPS with single-spin $A(\pm 1)$ matrices and $2\times 2$ block matrices $A(0,\ldots,15)$, 
with an without lattice symmetries.}
\label{fig4}
\vskip-3mm
\end{figure}

Fig.~\ref{fig4} shows the $D$ convergence of the energy and the magnetization of an $8\times 8$ lattice
close to the quantum-critical point; $h=3.044$. Results for comparison, $E/N=-3.23627(2)$ and $m^2=0.14073(1)$, 
were obtained with the stochastic series expansion (SSE) method \cite{sse}. Again it can be seen that the use of 
lattice symmetries is crucial for achieving good convergence. The convergence is also significantly 
better with the block matrices $A(0,\ldots,15)$ than with  single-spins matrices $A(\pm 1 )$.

\begin{figure}
\includegraphics[width=7.75cm, clip]{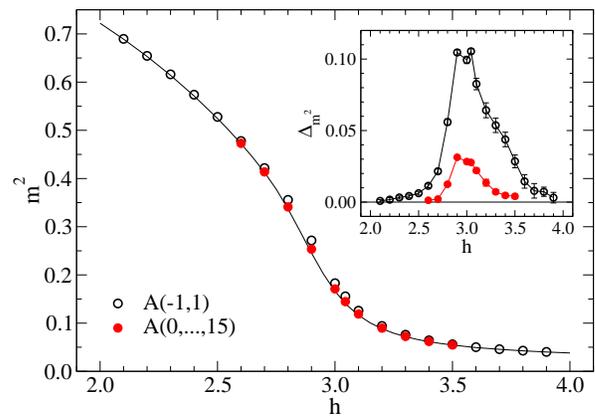}
\caption{(Color online) Squared magnetization as a function of the external field for an
$L=8$ system obtained with $D=4$ SR-MPS, using matrices either for single spins of for
blocks of $2\times 2$ spins at the lowest level. The solid curve shows results obtained with the 
approximation-free SSE method. Relative errors are shown in the inset.}
\label{fig5}
\vskip-3mm
\end{figure}

As with MPS or PEPS the accuracy of an SR-MPS calculation with fixed $D$ is expected to be lowest at a 
quantum-critical point. This is explicitly demonstrated for the SR-MPS description of the transverse-field
Ising model in Fig.~\ref{fig5}, using both single-spin and $2\times 2$ block matrices with $D=4$. 
With the single-spin matrices, the error in the squared magnetization in the neighborhood of the 
critical field is $\approx 10\%$, and with the $2\times 2$ block matrices it is $\approx 3\%$. 
The accuracy of the SR-MPS increases rapidly away from the critical point. 

In Ref.~\cite{string} string-state calculations for a $10\times 10$ lattice were reported. A magnetization curve exhibiting
a phase transition was obtained, but the accuracy is actually rather poor close to the critical point, with deviations of more 
than $50\%$ from the exact result and too little finite-size rounding. Since no systematic convergence tests were presented 
it is difficult to compare the performance of SR-MPS and string states directly.

Although the $8 \times 8$ lattice considered here is small in the context of QMC simulations of sign-problem-free models, 
the calculations demonstrate that the SR-MPS approach is a practically feasible. One important question is of course how
the $D$ required to obtain a desired accuracy grows as $N$ increases. In order for the scaling to be a power-law, instead 
of exponential, it is believed that an area law for the entanglement entropy has to be obeyed \cite{verst2}. Because of the 
lattice symmetries incorporated, the SR-MPS may satisfy such a law \cite{vidalcomment}. This, however, would still 
not guarantee a power-law scaling. Numerically it is also currently difficult to establish the scaling, because a range
of system sizes are needed. Calculations for $L=16$ at $h=h_{\rm c}$ give an energy error $\Delta_E<0.2\%$ for $D=5$ . 
Thus it is at least clear that one can access practically useful lattice sizes.

The SR-MPS scheme also work for frustrated systems. Preliminary calculations for an $L=8$ square-lattice $S=1/2$ 
Heisenberg model with a ratio $J_2/J_1=0.5$ of the second-nearest to nearest-neighbor interaction show a slower 
convergence with $D$ than in Fig.~\ref{fig4}, but it does appear feasible to reach the ground state.

It was shown here that incorporation of lattice symmetries in the wave function is crucial for achieving good convergence.
This has also been noted for one dimensional MPS \cite{porras}. Although the scaling of the computation increases by a factor 
of $N$, to $\propto D^3 N^2\ln(N)$ operations for updating the whole system, this can be partially alleviated by parallelizing 
the calculation of the spin-flip probability---the $8N$ different traces can be calculated
completely independently of each other. 

I would like to thank Y.-J. Kao and G. Vidal for stimulating discussions. This work was supported by the NSF under 
grant No.~DMR-0513930. Financial support from the National Center for Theoretical Sciences, Hsinchu, Taiwan, is
also gratefully acknowledged.

\null


\vskip-8mm
\begin{thebibliography}{00}

\bibitem{wilson}
K. G. Wilson, Rev. Mod. Phys. {\bf 47}, 773 (1995).

\bibitem{white}
S. R. White, Phys. Rev. Lett. {\bf 69}, 2863 (1992).

\bibitem{schollwock}
U. Schollw\"ock, Rev. Mod. Phys. {\bf 77}, 259 (2005).

\bibitem{ostlund}
S. \"Ostlund and S. Rommer, Phys. Rev. Lett. {\bf 75}, 3537 (1995).

\bibitem{nishino}
T. Nishino {\it et al.}, Nucl. Phys. B {\bf 575}, 504 (2000).

\bibitem{peps}
F. Verstraete and J. I. Cirac, Arxiv:cond-mat/0407066.

\bibitem{string}
N. Schuch, M. M. Wolf, F. Verstraete, and J. I. Cirac, ArXiv:0708.1567.

\bibitem{vidal}
G. Vidal, ArXiv:cond-mat/0512165.

\bibitem{awsandvidal}
A. W. Sandvik and G. Vidal, ArXiv:0708.2232 (to appear in Phys. Rev. Lett.).

\bibitem{verst}
F. Verstraete, D. Porras, J. I. Cirac, Phys. Rev. Lett. {\bf 93}, 227205 (2004).

\bibitem{ladders}
S. R. White and D. J. Scalapino, Phys. Rev. Lett. {\bf 91}, 136403 (2003).

\bibitem{liang}
S. Liang and H. Pang, Phys. Rev. B {\bf 49}, 9214 (1994).

\bibitem{vidal1}
G. Vidal, J. I. Latorre, E. Rico, and A. Kitaev, Phys. Rev. Lett. {\bf 90}, 227902 (2003). 

\bibitem{kadanov}
L. P. Kadanov, Physics {\bf 2}, 263 (1966).

\bibitem{hcrit}
H. Rieger and N. Kawashima, Eur. Phys. J. B {\bf 9}, 233 (1999).

\bibitem{sse}
A. W. Sandvik, Phys. Rev. E {\bf 68}, 056701 (2003). 

\bibitem{verst2}
F. Verstraete, M. M. Wolf, D. Perez-Garcia, and J. I. Cirac, Phys. Rev. Lett. 
{\bf 96}, 220601 (2006).

\bibitem{vidalcomment}
G. Vidal (private communication).

\bibitem{porras}
D. Porras, F. Verstraete, and J. I. Cirac, Phys. Rev. B {\bf 73}, 014410 (2006).
\end{thebibliography}
\end{document}